\documentclass[a4paper,11pt]{article}

\usepackage[latin1]{inputenc}

\usepackage[T1]{fontenc}
\usepackage[dvips]{graphicx}
\usepackage{times}

\usepackage{stmaryrd}

\usepackage{pstricks}
\usepackage{calrsfs}
\usepackage{subeqn}

\usepackage{pifont}
\usepackage{oldgerm}
\usepackage{color} 
\usepackage{amssymb,amsbsy}
\usepackage{epic,eepic}
\usepackage{texdraw}
\usepackage{dsfont}
\usepackage{latexsym}

\usepackage{accents}

\begin{document}

\title{Symmetry algebra of discrete KdV equations and corresponding differential-difference equations of Volterra type}
\author{Pavlos Xenitidis \\ 
School of Mathematics, University of Leeds, UK \\
P.Xenitidis@leeds.ac.uk}
\date{}
\maketitle

\begin{abstract}
A sequence of canonical conservation laws for all the Adler-Bobenko-Suris equations is derived and is employed in the construction of a hierarchy of master symmetries for equations H1-H3, Q1-Q3. For the discrete potential and Schwarzian KdV equations it is shown that their local generalized symmetries and non-local master symmetries in each lattice direction form centerless Virasoro type algebras. In particular, for the discrete potential KdV, the structure of its symmetry algebra is explicitly given. Interpreting the hierarchies of symmetries of equations H1-H3, Q1-Q3 as differential-difference equations of Yamilov's discretization of Krichever-Novikov equation, corresponding hierarchies of isospectral and non-isospectral zero curvature representations are derived for all of them. 
\end{abstract}

\section{Introduction}

Infinite hierarchies of symmetries and conservation laws are among the characteristics which most of the well known integrable systems possess \cite{M,O,JPWPhd,Z}. Their existence, serving as a definition of integrability, has been used in the classification of integrable differential equations, eg. \cite{ASY,M,JPWPhd}, and differential-difference equations \cite{ASY,LY1997, SY, Y1983, Y2006}, while this approach was recently extended to the case of difference equations \cite{LY2009}. An alternative approach to this notion of integrability is the existence of either a recursion operator \cite{O} or master symmetries \cite{F,O}, which can be used recursively for the construction of symmetries and conservation laws.

The above integrability criteria can be applied to difference equations as well, for which there exist also some other, alternative criteria to establish integrability \cite{GHRV}. Among them can be found the multidimensional consistency \cite{BS,NW,N2002} and the algebraic entropy \cite{BV}. Adler, Bobenko and Suris classified all the affine linear quadrilateral equations which possess the former property \cite{ABS,ABS1}, while it was shown in \cite{V} that the algebraic entropy of any equation in this class vanishes. Moreover, the symmetry analysis of these equations \cite{RHs, TTX} revealed that one of their symmetries plays the role of master symmetry with the help of which a hierarchy of symmetries was explicitly constructed in \cite{X}. Moreover, a recursion operator was recently found for the most general equation in this class along with some canonical conservation laws \cite{mwx,mwx1}, while some other conservation laws were found constructively in \cite{RHcl} and by implementation of the Gardner method in \cite{RS}. In this paper, we focus on the Adler-Bobenko-Suris (ABS) equations and explore further their integrability features. Among these integrability features are an infinite hierarchy of canonical conservation laws and a hierarchy of master symmetries for equations H1-H3 and Q1-Q3.

More precisely, first we present a derivation of a hierarchy of canonical conservation laws for all the ABS equations. This derivation is similar to the approach used in \cite{FF} and it is based on the use of the first canonical conservation law, found in \cite{mwx}, and the first master symmetry, given in \cite{X}. In particular, we introduce a potential in terms of which the conservation law can be given as the compatibility condition of a system of two, first order difference equations. Extending the action of the master symmetry to this potential results to the introduction of a new potential via another system of first order difference equations. The compatibility condition of the latter system is the next, higher order, canonical conservation law of the original ABS equation. Repeating the same argument with the new system, one is led to the next canonical conservation law, and, successively, a hierarchy of canonical conservation laws can be constructed.

The introduction of these potentials allows us to write in a simple and elegant form the higher order master symmetries of the ABS equations H1-H3 and Q1-Q3. More precisely, while the first master symmetry is local, the second member of this chain, and subsequently all the higher order members, depend explicitly on the potentials of canonical conservation laws. In spite of the fact that all the higher order master symmetries are nonlocal, it is proven that their commutators with the local generalized symmetries result to local generalized symmetries. This is in accordance with the continuous case \cite{S} where sufficient conditions guarantee the locality of these commutators. Actually, there exist obvious similarities in the form of the master symmetries constructed here and their forms given in Proposition 2 of \cite{S}.

For two of the ABS equations, namely equations H1 and Q1$_0$, we find the algebra of symmetries and master symmetries in each lattice direction. Specifically, we find explicit expressions for the characteristics of all symmetries, as well as how they act on the potentials of canonical conservation laws. Subsequently, we compute the commutators among the members of each hierarchy and find that they form centerless Virasoro type algebras. In particular for H1, we are able to find the commutators among the two sub-algebras and, thus, to describe its symmetry algebra.

From the interpretation of symmetries of the ABS equations as differential-difference equations \cite{LPSY, mwx1} follows that they correspond to particular cases of Yamilov's discrete Krichever-Novikov (YdKN) equation, or equation V4 with $\nu = 0$ in Yamilov's terminology \cite{Y2006}. This equation belongs to the class of Volterra type equations and it possesses a local master symmetry \cite{Y2006} which, in this context, is identified with the first master symmetry of the corresponding ABS equation. Here, we present hierarchies of such differential-difference equations and their corresponding hierarchies of master symmetries. Moreover, it is shown that the former possess an isospectral zero curvature representation while the latter possess the corresponding non-isospectral one \cite{FM,MF} and these representations are explicitly constructed.

The organization of the paper is as follows. The next section introduces our notation and reviews some basic properties of the ABS equations and their symmetry analysis. Section \ref{cl-ms} deals with the construction of conservation laws using the first master symmetry while Section \ref{secMS} contains the higher order master symmetries for equations H1-H3 and Q1-Q3. The derivation of the sub-algebra of symmetries in the $n$ direction for equations H1 and Q1$_0$ is the subject of Section \ref{secH1Q1}, while the symmetry algebra of H1 is studied in Section \ref{symH1}. The relation of symmetries to YdKN equation is explored in Section \ref{symdd} where also isospectral and non-isospectral hierarchies are explicitly constructed for these particular cases of YdKN equation. We conclude with an overall evaluation of the results obtained in the main body of the paper and several perspectives are discussed. In the Appendix, the ABS equations are given along with the basic polynomials related to them.

\section{The Adler-Bobenko-Suris equations} \label{ABS-not}

In this section we introduce our notation and present some properties of the ABS equations. Finally, we review some useful results on the symmetries and conservation laws of the latter equations. Further details about the ABS equations, their derivation and properties can be found in \cite{ABS,ABS1}, their symmetries in \cite{mwx,mwx1,RHs,TTX,X} and their conservation laws in \cite{mwx,RHcl}.

In what follows we will deal with scalar partial difference equations which involve the values of a function $u : {\mathds{Z}}^2 \mapsto {\mathds{R}}$ of two independent variables $n$ and $m$. The dependence of $u$ on $n$ and $m$ will be denoted by
$$u_{n+i,m+j}\,:=\,u(n+i,m+j)\,,\quad i,\,j\, \in\, {\mathds{Z}}\,. $$
In certain cases and in order to avoid the use of many indices, functions which depend explicitly or implicitly on $u$ and its shifts, i.e. functions of the form $G(n,m,u_{n+i_1,m+j_1},\ldots,u_{n+i_k,m+j_\ell})$, will be denoted simply by a letter without indices. In any event, it will be clear from the text if a symbol with no indices depends on the independent variables.

Also, we will use the two shifts operators and the two difference operators corresponding to the two lattice directions. The former will be denoted by ${\cal{S}}_n$ and ${\cal{S}}_m$, respectively, while the latter are defined as
$$\Delta_n\,:=\,{\cal{S}}_n\,-\,1 \quad {\mbox{and}} \quad  \Delta_m\,:=\,{\cal{S}}_m\,-\,1\,,$$
respectively. Finally, the variational derivative of a function $g(u_{n+i_1,m+j_1},\ldots,u_{n+i_k,m+j_\ell})$ is defined as
$$\frac{\delta g}{\delta u}\,\equiv\,\delta_u g\,:=\,\sum_{p \,=\, i_1}^{i_k} \sum_{q\,=\,j_1}^{j_\ell} {\cal{S}}^{\,-p}_n {\cal{S}}^{\,-q}_m \,\frac{\partial \,g}{\partial u_{n+p,m+q}}\,.$$

The ABS equations \cite{ABS,ABS1}, to which we are interested in, involve the values of $u$ at four different points which form an elementary quadrilateral on the lattice and depend explicitly on two lattice parameters $\alpha$ and $\beta$, i.e. they are equations of the form
\begin{equation}
Q_{n,m} \,:=\,Q(u_{n,m},u_{n+1,m},u_{n,m+1},u_{n+1,m+1},\alpha,\beta)\,=\,0\,. \label{ABS}
\end{equation}
The function $Q_{n,m}$ is an affine linear polynomial with respect to its first four arguments and equation (\ref{ABS}) is invariant under the mutual interchange of $u_{n+1,m}$ with $u_{n,m+1}$ accompanied with the interchange of the lattice parameters :
$$ Q(u_{n,m},u_{n,m+1},u_{n+1,m},u_{n+1,m+1},\beta,\alpha)\,=\,\pm\, Q(u_{n,m},u_{n+1,m},u_{n,m+1},u_{n+1,m+1},\alpha,\beta)\,.$$
In terms of the polynomial $Q_{n,m}$ and the discriminant-type operations
$$ {\cal{D}}_{x,y}\, g(x,y) \,:=\, g\, \partial_x\partial_y g\, -\, \left(\partial_x g\right) \left(\partial_y g\right) \,,\quad {\cal{D}}_{x}\, g(x,y) \,:=\, \left(\partial_x g\right)^2\, -\, 2\, g\, \partial_x^2 g\,, $$
one can define four functions. Namely,
\begin{subequations} \label{defpol}
\begin{eqnarray}
&& \kappa(\alpha,\beta)\,f(u_{n,m},u_{n+1,m},\alpha)\,\equiv\,\kappa(\alpha,\beta) f_{n,m}\,:=\, {\cal{D}}_{u_{n,m+1},u_{n+1,m+1}} Q_{n,m} \,, \label{deff}\\
&& h(u_{n+1,m},u_{n,m+1},\alpha,\beta)\,:=\,{\cal{D}}_{u_{n,m},u_{n+1,m+1}} Q_{n,m} \,,\label{defh}\\
&& r(u_{n,m})\,\equiv\,r_{n,m}\,:=\,{\cal{D}}_{u_{n+1,m}} f_{n,m}\,.\label{defr}
\end{eqnarray}
\end{subequations}
Here, $\kappa(\alpha,\beta)$ is an antisymmetric function of the lattice parameters, $f_{n,m}$ and $h$ are symmetric and biquadratic polynomials of the indicated values of $u$ and $r_{n,m}$ is, in general, a quartic polynomial. To make the paper self-contained, the ABS equations along with their corresponding polynomials are given in the Appendix.

All of the ABS equations possess a generalized symmetry \cite{TTX} of the form
\begin{equation}
V_1\,=\,\left(\frac{f_{n,m}}{w_{n,m}}\,-\,\frac{1}{2}\partial_{u_{n+1,m}} f_{n,m}\right)\,\partial_{u_{n,m}}\,=:\,R_1\,\partial_{u_{n,m}}\,, \label{V1}
\end{equation}
where
\begin{equation}
w_{n,m}\,:=\,u_{n+1,m}\,-\,u_{n-1,m}\,. \label{defw}
\end{equation}
Additionally, they possess an extended generalized symmetry \cite{TTX} 
\begin{equation}
G_1\,=\,n\,R_1\,\partial_{u_{n,m}}+\xi_1(\alpha) \partial_\alpha\,.\label{G1}
\end{equation}
The function $\xi_1$ appearing in $G_1$ is given in the following table.

\begin{center}
\begin{tabular}{l|ccccc}
& H1, H2 & H3 & Q1, Q2 & Q3 & Q4 \\ \hline
$\xi_1(\alpha)$ & $-1$ & $\frac{\alpha}{2}$ & $-1$ & $\frac{\alpha}{2}$ &$\frac{1}{2}\sqrt{4 \alpha^3- g_2 \alpha - g_3}$  
\end{tabular}
\end{center}

\noindent {\sl{Remark}} \hspace{.15cm} The form of function $\xi_1$ depends on the particular equation and its parametrization. Here we use the parametrization given in \cite{ABS} but always one can choose a parametrization in which $\xi_1 = 1$. In particular for Q4, the parametrization used here is due to Adler \cite{A} but one can choose to work in Hietarinta's parametrization \cite{H} at which $\xi_1 = 1$. \hfill $\Box$\\

One can construct a hierarchy of symmetries \cite{X} by employing the fact that $G_1$ is a master symmetry of $V_1$ \cite{RHs}, i.e. $[G_1,V_1] \ne 0$ and $[[G_1,V_1],V_1]=0$ \cite{F}. The resulting hierarchy has the following form \cite{X}
\begin{subequations} \label{VkR}
\begin{equation}
V_k\,:=\,\left[G_1,V_{k-1} \right]\,=\,R_k\,\partial_{u_{n,m}}\,\equiv\,{\cal{R}}^{\,k-1}\left(R_1\right)\,\partial_{u_{n,m}}\,,\quad k=2,3,\ldots, \label{Vk}
\end{equation} 
where
\begin{equation}
{\cal{R}}\,:=\,\sum_{p=-\infty}^{\infty} p \, \left({\cal{S}}_n^p R_1 \right) \partial_{u_{n+p,m}}\,+\,\xi_1(\alpha)\,\partial_\alpha\,. \label{R}
\end{equation}
\end{subequations}
From the above definition it is apparent that each $R_k$ depends on $(u_{n-k,m},u_{n-k+1,m},\ldots,u_{n+k-1,m},u_{n+k,m})$.

Due to the covariance of the ABS equations, a similar hierarchy of symmetries $\tilde{V}_\ell$ and a master symmetry $\tilde{G}_1$ follows from the interchanges $(n,u_{n+i,m+j},\alpha,\beta)\longrightarrow (m,u_{n+j,m+i},\beta,\alpha)$. It was shown in \cite{X} that the members of each hierarchy commute with each other, i.e. 
$$[V_i,V_j]\,=\, [\tilde{V}_i,\tilde{V}_j] \,=\,0\,,\quad \forall\, i,\,j\, \in\, {\mathds{N}}\,.$$
Moreover, the commutators between members of the two hierarchies and $G_1$, $\tilde{G}_1$ lead to trivial symmetries, i.e. their characteristics are identically zero on solutions of the corresponding equation $Q_{n,m}=0$ \cite{O}, implying that
$$[V_i,\tilde{V}_j] \,=\,[G_1,\tilde{V}_j]\,=\, [\tilde{G}_1,V_j] \,=\, [G_1,\tilde{G}_1]\,=\,0\quad {\mbox{on solutions of }}\, Q_{n,m}=0\,,\quad \forall\, i,\,j\, \in\, {\mathds{N}}\,.$$
In what follows, we will always assume that commutators of symmetries are evaluated on solutions of $Q_{n,m}=0$ without writing it out explicitly.

Another way to construct the above hierarchies  of symmetries is to use the pseudo-difference recursion operator given in \cite{mwx}. From the latter, one may also construct an infinite sequence of canonical conservation laws by considering the residues of this operator. The first conservation law, following from the logarithmic residue of this pseudo-difference operator, is
\begin{equation}
 \Delta_m \left(\log\frac{w_{n,m}^2}{f_{n-1,m}}\right)\,=\,\Delta_n\left(\log \frac{h(u_{n-1,m},u_{n,m+1},\alpha,\beta)}{\kappa(\alpha,\beta)f_{n-1,m}}\right)\,.\label{cl0}
\end{equation}
The form of a conservation law implies that one can introduce, at least locally, a potential function and represent the conservation law as the compatibility condition of a first order system. For the above conservation law, we introduce the potential $\phi^{(0)} \equiv \phi^{(0)}_{n,m}$ through the system of equations 
\begin{equation}
\Delta_n \phi^{(0)}\,=\,\rho^{(0)}_{n,m}:=\log\frac{w_{n,m}^2}{f_{n-1,m}}\,,\quad \Delta_m \phi^{(0)}\,=\,\sigma_{n,m}^{(0)}:=\log \frac{h(u_{n-1,m},u_{n,m+1},\alpha,\beta)}{\kappa(\alpha,\beta)f_{n-1,m}}\,,\label{ABS-sys-0-cl}
\end{equation}
the compatibility condition of which is relation (\ref{cl0}). In what follows, we will refer to system (\ref{ABS-sys-0-cl}) and other similar systems, like (\ref{cl-1}) and (\ref{cl-2}) below, as conservation laws as well.

\section{Conservation laws and master symmetries} \label{cl-ms}

It is well known \cite{FF,F,OZF,O} that one can employ symmetries and master symmetries in the construction of conservation laws. In this section we follow an approach similar to this in \cite{FF} which makes use of master symmetry $G_1$ and first canonical conservation law (\ref{ABS-sys-0-cl}) in terms of which an infinite sequence of conservation laws for all of the ABS equations is derived.

We start with the system constituted by the ABS equation (\ref{ABS}) and system (\ref{ABS-sys-0-cl}) for the potential $\phi^{(0)}$. This system is consistent and its symmetries obviously follow from the corresponding ones of the ABS equation by extending them to the $\phi^{(0)}$-direction. This means that a generalized symmetry of this system has the form
\begin{equation}
V_1\,=\,R_1 \partial_{u_{n,m}} + \Phi_{1}^{(0)} \partial_{\phi^{(0)}}\,,
\end{equation}
where $\Phi_{1}^{(0)}$ is a function to be determined so that
\begin{subequations} \label{det12}
\begin{eqnarray}
\Delta_n\Phi_{1}^{(0)} = {\cal{S}}_n(R_{1}) \partial_{u_{n+1,m}} \rho_{n,m}^{(0)} + R_{1} \partial_{u_{n,m}} \rho_{n,m}^{(0)} + {\cal{S}}_n^{-1}(R_{1}) \partial_{u_{n-1,m}} \rho_{n,m}^{(0)}\,,\label{det1-0}\\
\Delta_m \Phi_{1}^{(0)} = {\cal{S}}_m(R_{1}) \partial_{u_{n,m+1}} \sigma_{n,m}^{(0)} + R_{1} \partial_{u_{n,m}} \sigma_{n,m}^{(0)} + {\cal{S}}_n^{-1}(R_{1}) \partial_{u_{n-1,m}} \sigma_{n,m}^{(0)}\,.\label{det2-0}
\end{eqnarray}
\end{subequations}

Using the identity
\begin{equation}
\frac{f_{n,m}}{w_{n,m}}-\frac{1}{2} \partial_{u_{n+1,m}}f_{n,m} = \frac{f_{n-1,m}}{w_{n,m}}+\frac{1}{2} \partial_{u_{n-1,m}}f_{n-1,m}\,,
\end{equation}
which holds for any symmetric and bi-quadratic polynomial $f$, it is a straightforward computation to show that the right hand side of equation (\ref{det1-0}) can be written as 
$\Delta_n \left(R_1 \partial_{u_{n,m}} \rho_{n-1,m}^{(0)} \right)$, which in turn implies that $\Phi_1^{(0)} \,=\,R_1 \partial_{u_{n,m}} \rho_{n-1,m}^{(0)} + c_m$, where $c_m$ is a function of $m$. 
Finally, in view of the latter relation and using equation (\ref{ABS}), the second of equations (\ref{det12})  implies that $c_m = c_0$, where $c_0$ is a constant. Since this parameter reflects the translation invariance $\phi^{(0)} \mapsto \phi^{(0)}\,+\,\epsilon$ of system (\ref{ABS-sys-0-cl}), we set $c_0=0$ and, hence, a symmetry generator of system (\ref{ABS}), (\ref{ABS-sys-0-cl}) has the form 
\begin{equation}
V_1 = R_1 \partial_{u_{n,m}} + R_1 \partial_{u_{n,m}} \rho_{n-1,m}^{(0)}  \partial_{\phi^{(0)}}\,.
\end{equation}

Applying the same procedure with master symmetry $G_1$ by extending it to the direction of potential $\phi^{(0)}$
\begin{equation}
G_1 \,=\,n\,R_1 \partial_{u_{n,m}} \,+\,\left(n R_1 \partial_{u_{n,m}} \rho_{n-1,m}^{(0)} + \phi^{(1)}\right)\partial_{\phi^{(0)}} + \xi_1(\alpha) \partial_\alpha\,,
\end{equation}
it follows that $\phi^{(1)}$ must obey the system of equations
\begin{equation}
\Delta_n \phi^{(1)}\,=\,\rho^{(1)}_{n,m}:= {\cal{R}} \left(\rho_{n,m}^{(0)}\right)- {\cal{S}}_n\Phi_1^{(0)}\,,\quad \Delta_m \phi^{(1)}\,=\,\sigma_{n,m}^{(1)}:={\cal{R}}(\sigma_{n,m}^{(0)})\,. \label{cl-1}
\end{equation}
In particular, using the identity 
\begin{equation}
\xi_1(\alpha) \,\partial_\alpha f_{n,m} \,-\,\frac{1}{2}\left|\begin{array}{cc} f_{n,m} & \partial_{u_{n,m}} f_{n,m} \\ \partial_{u_{n+1,m}} f_{n,m}  & \partial_{u_{n,m}} \partial_{u_{n+1,m}} f_{n,m} \end{array} \right|\,=\,0\,, \label{mscon}
\end{equation}
which holds for the polynomials $f_{n,m}$ corresponding to the ABS equations, function $\rho_{n,m}^{(1)}$ can be written also as
$$\rho_{n,m}^{(1)} =  \frac{2 R_1}{w_{n-1,m}} - \partial_{u_{n,m}} R_1\,.$$
The compatibility condition of equations (\ref{cl-1}), i.e. $\Delta_m \rho_{n,m}^{(1)} = \Delta_n \sigma_{n,m}^{(1)}$, is satisfied in view of the corresponding discrete equation. Thus, it is another conservation law of the ABS equation $Q_{n,m}=0$, which actually coincides with the second canonical conservation law given in \cite{mwx}.

Now we can include $\phi^{(1)}$ in our derivation by considering the system
$$Q_{n,m} = 0\,,\quad \Delta_n\phi^{(i)}=\rho^{(i)}_{n,m}\,,\quad \Delta_m\phi^{(i)}=\sigma^{(i)}_{n,m}\,,\quad i=0,1\,.$$
To find its symmetries we have to extend $V_1$ and $G_1$ in the direction of the new potential. These considerations result to
\begin{eqnarray}
V_1 &=&  R_1 \partial_{u_{n,m}} + R_1 \partial_{u_{n,m}} \rho_{n-1,m}^{(0)}  \partial_{\phi^{(0)}} \nonumber \\
& & + \left(R_1 \partial_{u_{n,m}} \rho_{n-1,m}^{(1)} + \frac{r_{n,m}}{2 w_{n-1,m}^2} - \frac{\partial_{u_{n,m}} r_{n,m}}{4 w_{n-1,m}} + \frac{\partial_{u_{n,m}}^2 r_{n,m}}{24}\right) \partial_{\phi^{(1)}}\,,
\end{eqnarray}
and
\begin{equation}
G_1 \,=\,n\,V_1 + \phi^{(1)} \partial_{\phi^{(0)}}  + \phi^{(2)} \partial_{\phi^{(1)}} + \xi_1(\alpha) \partial_\alpha\,,
\end{equation}
respectively, where $r_{n,m}$ is the discriminant of $f_{n,m}$ as a polynomial of $u_{n+1,m}$, cf. relation (\ref{defr}). Function $\phi^{(2)}$ appearing in $G_1$ obeys the system of equations
\begin{equation}
\Delta_n \phi^{(2)}\,=\, \rho^{(2)}_{n,m}:= {\cal{R}} (\rho_{n,m}^{(1)}) - {\cal{S}}_n\left(\Phi_1^{(1)}\right)\,, \quad  \Delta_m \phi^{(2)}\,=\,\sigma_{n,m}^{(2)}:={\cal{R}}(\sigma_{n,m}^{(1)})\,, \label{cl-2}
\end{equation}
where $\Phi_1^{(1)}$ is the $\partial_{\phi^{(1)}}$--component of $V_1$.
The compatibility condition $\Delta_m \rho_{n,m}^{(2)} = \Delta_n \sigma_{n,m}^{(2)} $ of the above system is satisfied by taking into account only equation $Q_{n,m}=0$ and its shifts, i.e. it is another conservation law of $Q_{n,m}=0$. Actually, it is equivalent\footnote{Two conserved densities are called equivalent if the variational derivative of their difference is identically zero \cite{mwx}.} to the third conservation law derived in \cite{mwx}.

We can continue this procedure by including $\phi^{(2)}$ in our system and extending the two symmetry generators $V_1$ and $G_1$ in this direction. This leads to the introduction of a new potential $\phi^{(3)}$ through the system
\begin{equation}
\Delta_n \phi^{(3)}\,=\, \rho^{(3)}_{n,m}:= {\cal{R}} (\rho_{n,m}^{(2)}) - {\cal{S}}_n\left(\Phi_1^{(2)}\right)\,, \quad  \Delta_m \phi^{(3)}\,=\,\sigma_{n,m}^{(3)}:={\cal{R}}(\sigma_{n,m}^{(2)})\,, \label{cl-3}
\end{equation}
where $\Phi_1^{(2)}$ is the $\partial_{\phi^{(2)}}$--component of $V_1$, i.e.
$$\Phi_1^{(2)} =  R_1 \partial_{u_{n,m}} \rho_{n-1,m}^{(2)} + {\cal{R}}\left(\frac{r_{n,m}}{2 w_{n-1,m}^2} - \frac{\partial_{u_{n,m}} r_{n,m}}{4 w_{n-1,m}} + \frac{\partial_{u_{n,m}}^2 r_{n,m}}{24}\right)\,.$$
Again, the compatibility condition $\Delta_m \rho_{n,m}^{(3)} = \Delta_n \sigma_{n,m}^{(3)}$ is another conservation law for equation (\ref{ABS}).

We can repeat infinitely many times this procedure and construct an infinite hierarchy of conservation laws along with the corresponding functions $\Phi_1^{(k)}$. In particular, the conserved densities and fluxes are given recursively by the relations
\begin{equation}
\rho_{n,m}^{(k)} \,:=\,{\cal{R}}\left(\rho_{n,m}^{(k-1)}\right) \,-\,{\cal{S}}_n\Phi_1^{(k-1)}\,,\quad \sigma_{n,m}^{(k)}\,=\,{\cal{R}}\left(\sigma_{n,m}^{(k-1)}\right)\,,\quad k=1,2,3,\cdots, \label{def-dens-flux}
\end{equation}
or, equivalently, 
\begin{equation}
\rho_{n,m}^{(k)} \,=\, {\cal{R}}^k\left(\rho_{n,m}^{(0)} \right) - \sum_{j=0}^{k-1} {\cal{R}}^j\left({\cal{S}}_n\left(\Phi_1^{(k-j-1)}\right)\right)\,,\quad \sigma_{n,m}^{(k)}\,=\,{\cal{R}}^k\left(\sigma_{n,m}^{(0)}\right)\,,\quad k=1,2,3,\cdots.
\end{equation}
 However, it is not easy to derive a simple recursive formula for functions $\Phi_1^{(k)}$. Here we list the first five which we were able to find explicitly.
\begin{eqnarray*}
&& \Phi_1^{(0)} = R_1 \partial_{u_{n,m}} \rho_{n-1,m}^{(0)}  \\
&& \Phi_1^{(1)} = R_1 \partial_{u_{n,m}} \rho_{n-1,m}^{(1)} + \frac{r_{n,m}}{2 w_{n-1,m}^2} - \frac{\partial_{u_{n,m}} r_{n,m}}{4 w_{n-1,m}} + \frac{\partial_{u_{n,m}}^2 r_{n,m}}{24}\\
&& \Phi_1^{(2)} = R_1 \partial_{u_{n,m}} \rho_{n-1,m}^{(2)} + {\cal{R}}\left(\frac{r_{n,m}}{2 w_{n-1,m}^2} - \frac{\partial_{u_{n,m}} r_{n,m}}{4 w_{n-1,m}} + \frac{\partial_{u_{n,m}}^2 r_{n,m}}{24}\right)  \\
&& \Phi_1^{(3)} =  R_1 \partial_{u_{n,m}} \rho_{n-1,m}^{(3)} + {\cal{R}}^{\,2}\left(\frac{r_{n,m}}{2 w_{n-1,m}^2} - \frac{\partial_{u_{n,m}} r_{n,m}}{4 w_{n-1,m}} + \frac{\partial_{u_{n,m}}^2 r_{n,m}}{24}\right) - \frac{r_{n,m} f_{n-2,m} f_{n-3,m}}{w_{n-1,m}^4 w_{n-2,m}^2}\\
&&\Phi_1^{(4)} =  R_1 \partial_{u_{n,m}} \rho_{n-1,m}^{(4)} + {\cal{R}}^{\,3}\left(\frac{r_{n,m}}{2 w_{n-1,m}^2} - \frac{\partial_{u_{n,m}} r_{n,m}}{4 w_{n-1,m}} + \frac{\partial_{u_{n,m}}^2 r_{n,m}}{24}\right) - 2 {\cal{R}}\left( \frac{r_{n,m} f_{n-2,m} f_{n-3,m}}{w_{n-1,m}^4 w_{n-2,m}^2}\right)
\end{eqnarray*}
Only for two equations, namely H1 and Q1$_{0}$, for which $r_{n,m} \equiv 0$, one can generate recursively the whole hierarchy of $\Phi_1^{(k)}$ and this construction will be given in Section \ref{secH1Q1}.

It is obvious that the conserved density $\rho_{n,m}^{(k)}$ depends only on $(u_{n-k-1,m},\ldots,u_{n,m},u_{n+1,m})$, while $\Phi_1^{(k)}$ depends on $(u_{n-k-2,m},u_{n-k-1,m},\ldots,u_{n,m},u_{n+1,m})$. We can prove inductively that this construction does not terminate. The first five conserved densities are neither trivial nor equivalent to each other, which can be checked directly using the above formulas and any of the ABS equations. Suppose that after $k$ iterations, density $\rho_{n,m}^{(k-1)}$ is neither trivial nor equivalent to any other lower order density, but the next one $\rho_{n,m}^{(k)}$ is equivalent to $\rho_{n,m}^{(\ell)}$, where $\ell < k$. This means that $\delta_u\left(\rho_{n,m}^{(k)}-\rho_{n,m}^{(\ell)}\right)=0$. But the only term in the difference of these two densities which involves the value $u_{n-k-1,m}$ is $-k \,{\cal{S}}_n^{-k}\left(R_1\right)\partial_{u_{n-k,m}} \rho_{n,m}^{(k-1)}\,\ne \,0$ and its contribution to the variational derivative is non-zero, since $\partial_{u_{n,m}}\left({\cal{S}}_n^{k+1}\left({\cal{S}}_n^{-k}\left(R_1\right)\partial_{u_{n-k,m}} \rho_{n,m}^{(k-1)}\right) \right)=f_{n+1,m} \left( \partial_{u_{n+1,m}} \rho_{n+k+1,m}^{(k-1)}\right)/w_{n+1,m}^2 \ne 0$. Thus, $\rho_{n,m}^{(k)}$ is not equivalent to any lower order density. From the same arguments, it follows that $\rho_{n,m}^{(k)}$ cannot be trivial as well.

\section{Hierarchies of master symmetries} \label{secMS}

Once we have constructed the hierarchy of potentials $\phi^{(k)}$ and the symmetry generators
\begin{eqnarray}
V_1 &=& R_1\partial_{u_{n,m}} \,+\,\sum_k \Phi_1^{(k)} \partial_{\phi^{(k)}} \,,\\
G_1 &=& n\,R_1 \partial_{u_{n,m}} \,+\, \sum_k \left(n\,\Phi_1^{(k)} + \phi^{(k+1)}\right)\partial_{\phi^{(k)}}\,+\,\xi_1(\alpha)\partial_\alpha\,,
\end{eqnarray}
we can find how the higher order symmetries $V_\ell$ (\ref{VkR}) act on the potentials $\phi^{(k)}$. This can be done by considering the commutators of $G_1$ with these symmetries starting with $V_1$. To this end, the symmetry hierarchy is given by
\begin{subequations}
\begin{equation}
V_\ell \,=\, R_\ell\, \partial_{u_{n,m}}\,+\,\sum_k \Phi_\ell^{(k)} \partial_{\phi^{(k)}}\,,\quad \ell=1,2,3,\ldots,
\end{equation}
where
\begin{equation}
 \Phi_{\ell}^{(k)}\,=\,\sum_{j=0}^{\ell-1} \left(\begin{array}{c} \ell-1 \\ j\end{array} \right) (-1)^{\ell+j-1} {\cal{R}}^{j}\left(\Phi_1^{(k+\ell-j-1)}\right)  \,,\quad \ell=1,2,3,\ldots,\,\,k=0,1,2,\ldots\,. \label{Fkl}
\end{equation}
\end{subequations}

Using the hierarchy of potentials $\phi^{(k)}$, one can derive a corresponding hierarchy of non-local master symmetries for equations H1--H3 and Q1--Q3. More precisely, this can be done starting with $G_1$ above and 
\begin{equation}
G_2 \,=\, \left(n R_2 \,- \,\phi^{(1)} R_1\,-\,\frac{r_{n,m}}{2\, w_{n-1,m}} \right)\partial_{u_{n,m}} \,+ \,\xi_2(\alpha)\,\partial_\alpha\,+\,\zeta_2(\alpha,\beta)\,\partial_\beta\,, \label{2ms}
\end{equation}
where functions $\xi_2(\alpha)$, $\zeta_2(\alpha,\beta)$ appearing in $G_2$ are given in the following list.
\begin{eqnarray*}
{\rm{H1}} &:& (\alpha-\beta) (\zeta_2-\xi_2) \,-\,1\,=\,0\quad \Longrightarrow \quad \xi_2 = 0\,, \quad \zeta_2 = \frac{1}{\alpha-\beta}\\
{\rm{H2}} &:& \xi_2 = 0\,, \quad \zeta_2 = \frac{1}{\alpha-\beta}\\
{\rm{H3}}_{0} &:&   4 (\alpha^2-\beta^2) (\alpha \zeta_2-\beta \xi_2) - \alpha \beta (\alpha^2+\beta^2) = 0 \quad \Longrightarrow \quad  \xi_2 = - \frac{\alpha}{4},\,\, \zeta_2 = \frac{\beta^3}{2 (  \alpha^2-\beta^2)} \\
{\rm{H3}}_{\delta } &:& \xi_2 = - \frac{\alpha}{4}\,,\quad \zeta_2 = \frac{\beta^3}{2 (  \alpha^2-\beta^2)}\\
{\rm{Q1}}_0 &:& \alpha \zeta_2-\beta \xi_2 - \beta \left(\frac{1}{\alpha}+\frac{1}{\alpha-\beta}\right) = 0  \quad\Longrightarrow \quad   \xi_2 =- \frac{1}{\alpha}\,,\quad \zeta_2 =  \frac{1}{\alpha-\beta}-\frac{1}{\alpha} \\ 
{\rm{Q1}}_\delta &:& \xi_2 =- \frac{1}{\alpha}\,,\quad \zeta_2 =  \frac{1}{\alpha-\beta}-\frac{1}{\alpha} \\
{\rm{Q2}} &:&\xi_2 =-\frac{1}{\alpha}\,,\quad \zeta_2 = \frac{1}{\alpha-\beta} - \frac{1}{\alpha} \\
{\rm{Q3}} &:& \xi_2 =  - \frac{\alpha (\alpha^2+1)}{4 (\alpha^2-1)}\,,\quad \zeta_2 = \frac{\alpha^2 \beta (\beta^2-1)}{2 (\alpha^2-1) (\alpha^2-\beta^2)} 
\end{eqnarray*}

\noindent {\emph{Remark}} \hspace{.15cm} In the case of H1, H3$_{0}$ and Q1$_0$, functions $\xi_2$, $\zeta_2$ are not completely determined and one has to choose a solution of the corresponding linear algebraic equations. In the list, a particular solution for each of these equations is given and this solution will be used in what follows. Our choices for H3$_0$ and Q1$_0$ coincide with the corresponding functions of their $\delta$-counterparts, while for H1 with the ones for H2. \hfill $\Box$\\

\noindent {\emph{Remark on the derivation of $G_2$}} \hspace{.15cm} For H1 the characteristic of $G_2$ was derived from the first master symmetry $G_1$ by applying to the latter the recursion operator given in \cite{mwx}. Then the linear algebraic equation for functions $\xi_2$ and $\zeta_2$ follows from the requirement the resulting function to be a symmetry of H1. The same approach was used for the remaining equations admitting first order recursion operators, namely equations H2, H3, Q1 and Q3$_0$. For Q2 and Q3$_\delta$, the form of $G_2$ was verified by direct calculations. \hfill $\Box$\\

One can verify that $G_2$ is another symmetry of the corresponding ABS equation in a straightforward manner. Specifically, it is sufficient to show that the infinitesimal criterion 
$$\sum_{i=0}^{1}\sum_{j=0}^1 {\cal{S}}_n^{\,i}{\cal{S}}_m^{\,j}\left(n R_2 \,- \,\phi^{(1)} R_1\,-\,\frac{r_{n,m}}{2 \,w_{n-1,m}}\right) \frac{\partial Q_{n,m}}{\partial u_{n+i,m+j}} + \xi_2 \partial_\alpha Q_{n,m} + \zeta_2 \partial_\beta Q_{n,m}= 0$$
holds on solutions of $Q_{n,m}=0$ and its shifts, and by taking into account that
$$\Delta_n \phi^{(1)} = \rho_{n,m}^{(1)}\,,\quad \Delta_m \phi^{(1)} = \sigma_{n,m}^{(1)}\,.$$
Moreover, the commutator of $G_2$ with any of the symmetries $V_j$ will always produce local symmetries. Indeed, writing out explicitly the commutator $[G_2,V_j]$ and using the facts that all symmetries commute $[V_i,V_j]=0$ and $\Delta_n \phi^{(1)} = \rho_{n,m}^{(1)}$, one finds that the characteristic of the resulting generator has the following form.
\begin{eqnarray*}
&& \sum_{k=-j}^{j} k\,\left({\cal{S}}_n^k R_2 \right) \frac{\partial R_j}{\partial u_{n+k,m}} - \xi_2(\alpha) \frac{\partial R_j}{\partial \alpha} + R_1 \Phi_j^{(1)} - \sum_{k=-j}^{j} \frac{r_{n+k,m}}{2 w_{n+k-1,m}}\,\frac{\partial R_j}{\partial u_{n+k,m}} +\\ 
&&\sum_{k=-2}^{0} \left({\cal{S}}_n^k R_j \right) \partial_{u_{n+k,m}}\left( \frac{r_{n,m}}{2 w_{n-1,m}}\right) - \sum_{k=1}^{j} \left( \sum_{i=0}^{k-1} {\cal{S}}_n^i \rho_{n,m}^{(1)} \right)\frac{\partial R_j}{\partial u_{n+k,m}} + \sum_{k=-j}^{-1} \left( \sum_{i=k}^{-1} {\cal{S}}_n^i \rho_{n,m}^{(1)} \right) \frac{\partial R_j}{\partial u_{n+k,m}}.
\end{eqnarray*}
All the functions involved in the above expression are local, i.e. do not depend on the potentials $\phi^{(i)}$, implying that indeed $[G_2,V_j]$ is local. Moreover, it does not depend explicitly on $n$.

Additionally, $G_2$ is another master symmetry for $V_1$ since it can be shown that
\begin{eqnarray*}
{\mbox{for H1, H2, Q1 and Q2}} &:& [G_2,V_1] \,=\,\frac{1}{2} V_3 \,,\\
{\mbox{for H3 and Q3}} &:& [G_2,V_1] \,=\,\frac{1}{2} V_3 + \frac{1}{12} V_1\,,
\end{eqnarray*}
implying in each case that $[[G_2,V_1],V_1]=0$. Finally, it can be verified that the action of $G_2$ on conservation law (\ref{cl0}) leads to the second conservation law (\ref{cl-2}), i.e. $G_2$ produces higher canonical conservation laws in the same fashion as $G_1$ does.

One can construct the next master symmetry by considering the commutator of $G_1$ and $G_2$. At this point it is more convenient to consider separately equations H1, H2, Q1 and Q2 from equations H3 and Q3. For the former equations one finds that the higher order master symmetries have the following forms.
\begin{eqnarray}
&& G_{\ell}\,:=\,\left[G_1,G_{\ell-1}\right]\,=\, \left(\frac{n}{\ell-1} R_\ell - \sum_{j=0}^{\ell-2} \left(\begin{array}{c} \ell-2 \\ j\end{array} \right) R_{j+1} \phi^{(\ell-j-1)} - {\cal{R}}^{\,\ell-2}\left(\frac{r_{n,m}}{2 \, w_{n-1,m}}\right)\right)\partial_{u_{n,m}} \nonumber \\
&&\phantom{ G_{\ell}\,:=\,\left[G_1,G_{\ell-1}\right]\,=\,}\,\, +  [\xi_1(\alpha)\, \partial_\alpha\,,\, \xi_{\ell-1}(\alpha) \, \partial_\alpha \, +\, \zeta_{\ell-1}(\alpha,\beta)\, \partial_\beta] \,,\quad \ell = 3,4,5,\cdots\,. \label{H12Q12Gl}
\end{eqnarray}

For equations H3 and Q3, the resulting symmetries partly follow this pattern. More precisely, the next master symmetry $G_3$, following from the commutator $[G_1,G_2]$, reads as follows
\begin{subequations} \label{msH3Q3}
\begin{equation}
G_3= \left(\frac{n}{2} R_3 - \frac{n}{12} R_1 - \phi^{(1)} R_2 - \phi^{(2)} R_1 - {\cal{R}} \left(\frac{r_{n,m}}{2 \, w_{n-1,m}}\right)\right)\partial_{u_{n,m}}+ \left[\,\xi_1 \,\partial_\alpha\,,\,\xi_2\, \partial_\alpha\,+\,\zeta_2 \,\partial_\beta \,\right] \,, \label{G3HQ3}
\end{equation}
while the next two symmetries $G_4$ and $G_5$ follow the general pattern (\ref{H12Q12Gl}). Continuing with the next two symmetries, one finds that
\begin{equation}
G_6 = \left(\frac{n}{5} R_6 - \frac{n}{30} R_2 - \sum_{j=0}^{4} \left(\begin{array}{c} 4 \\ j\end{array} \right) \phi^{(5-j)} R_{j+1} - {\cal{R}}^{\,4} \left(\frac{r_{n,m}}{2 \, w_{n-1,m}}\right)\right)\partial_{u_{n,m}} \,+ \,\left [\xi_1 \partial_\alpha,\xi_5 \partial \alpha+ \zeta_5 \partial_\beta\right]
\end{equation}
and
\begin{equation}
G_7 = \left(\frac{n}{6} R_7 - \frac{n}{12} R_3 - \sum_{j=0}^{5} \left(\begin{array}{c} 5 \\ j\end{array} \right) \phi^{(6-j)} R_{j+1} - {\cal{R}}^{\,5} \left(\frac{r_{n,m}}{2 \, w_{n-1,m}}\right)\right)\partial_{u_{n,m}} \,+\, \left [\xi_1 \partial_\alpha,\xi_6 \partial \alpha+ \zeta_6 \partial_\beta\right]\,,
\end{equation}
\end{subequations}
respectively.

It can be shown by induction that the commutators $[G_i,V_j]$ result to local symmetries. As we have shown, this is the case when $i=1$ and $i=2$. Using the definition of $G_3$ and the Jacobi identity, the commutator $[G_3,V_j]$ can be written as $[G_1,[G_2,V_j]]- [G_{2},V_{j+1}]$. Since $G_1$ is local and we have shown that $[G_2,V_i]$ is local for any $i$, it follows that $[G_3,V_j]$ is local as well. Assuming that $[G_k,V_j]$ is local for some $k>3$ and for all $j$ and using the same arguments, it follows that $[G_{k+1},V_j] = [G_1,[G_k,V_j]]- [G_{k},V_{j+1}]$ is also local.\\

\noindent {\sl{Remark}}\hspace{.15cm} In the continuous case there exist sufficient conditions which guarantee the locality of symmetries resulting from the commutator of master symmetries with other symmetries \cite{S}. More precisely, Proposition 2 in \cite{S} gives the necessary conditions on the form of the non-local part of a master symmetry so that to produce local symmetries. Actually, comparing the non-local part of all the above master symmetries with the form assumed in Proposition 2 of \cite{S}, it is obvious that $G_k$'s are of this form, with $W_\gamma$ and $\omega_\gamma$ in \cite{S} identified respectively with $R_\gamma$ and $\phi^{(\gamma)}$ here. \hfill $\Box$ \\

Once the symmetry sub-algebra $\mathfrak{n}$ generated by $V_\ell$ and $G_\ell$ has been constructed, another symmetry sub-algebra $\mathfrak{m}$ follows from the covariance of the ABS equations. Specifically, hierarchies of symmetries $\tilde{V}_\ell$, master symmetries $\tilde{G}_\ell$ and potentials $\tilde{\phi}^{(\ell)}$ follow from the usual interchanges $(n,u_{n+i,m+j},\alpha,\beta)\longrightarrow (m,u_{n+j,m+i},\beta,\alpha)$ which generate the symmetry sub-algebra $\mathfrak{m}$. These two sub-algebras, along with the point symmetries each equation possesses, form the symmetry algebra $\mathfrak{h}$ of the corresponding equation. To find the commutators among the elements of the two sub-algebras, one has to extend the action of $V_k$ and $G_k$ in the direction of the potentials $\tilde{\phi}^{(i)}$ and then evaluate the commutators. However, to find general expressions valid for any equation in this class is very difficult. This can be done for each equation separately and for equation H1 the corresponding results are presented in Section \ref{symH1}.

\section{Discrete potential and Schwarzian KdV equations} \label{secH1Q1}

Having constructed a sequence of symmetries and master symmetries in the manner described in the previous sections, one interesting question is what is the algebraic structure of these symmetries. In the case of differential and differential-difference equations usually one finds that all these form centerless Virasoro type algebras. In this section we show that the symmetries $(V_k,G_k)$ of H1, as well as of Q1 with $\delta=0$, form a centerless Virasoro type algebra. This is achieved by finding explicit expressions for all symmetry generators and their action on the potentials $\phi^{(k)}$, and, consequently, computing their commutators. Due to the covariance of H1 and Q1$_0$, similar relations hold for the symmetries $(\tilde{V}_k,\tilde{G}_k)$ in the $m$ direction.

The lattice potential KdV equation (also known as H1)
\begin{equation}
(u_{n,m}-u_{n+1,m+1}) (u_{n+1,m}-u_{n,m+1}) - \alpha + \beta = 0 \label{H1}
\end{equation}
and the lattice Schwarzian KdV equation (also referred to as Q1$_{0}$)
\begin{equation}
\alpha (u_{n,m}-u_{n,m+1}) (u_{n+1,m}-u_{n+1,m+1}) - \beta (u_{n,m}-u_{n+1,m}) (u_{n,m+1}-u_{n+1,m+1})= 0 \label{Q10}
\end{equation}
are the only equations in the ABS classification \cite{ABS} for which the polynomial $r_{n,m}$ is identically zero. For these two equations, it is easier to construct the hierarchies of conservation laws and hierarchies of symmetries using the results of the previous sections.

The characteristic $R_1$ of the first symmetry $V_1$ is
\begin{subequations}
\begin{equation}
{\mbox{for H1 :}} \,\,R_1\,=\,\frac{1}{w_{n,m}}\,,
\end{equation}
and
\begin{equation}
 {\mbox{for Q1}}_{0} \,:\,  R_1\,=\,\frac{(u_{n+1,m}-u_{n,m}) (u_{n-1,m}-u_{n,m})}{\alpha\,w_{n,m}}\,,
\end{equation}
\end{subequations}
respectively. Now, starting with the conservation law (\ref{ABS-sys-0-cl}) which takes the following form
\begin{subequations}
\begin{equation}
{\mbox{for H1}}\,\,:\,\, \rho_{n,m}^{(0)}=2 \log \left(u_{n+1,m}-u_{n-1,m}\right)\,,\quad \sigma_{n,m}^{(0)} = \log\frac{(u_{n-1,m}-u_{n,m+1})^2}{\beta-\alpha}\,,
\end{equation}
and
\begin{equation}
{\mbox{for Q1}}_{0} \,\,:\,\,  \rho_{n,m}^{(0)}=2 \log \frac{u_{n+1,m}-u_{n-1,m}}{u_{n,m}-u_{n-1,m}} + \log \alpha ,\, \sigma_{n,m}^{(0)}=2 \log\frac{u_{n-1,m}-u_{n,m+1}}{u_{n,m}-u_{n-1,m}}+ \log\frac{\alpha}{\beta-\alpha},
\end{equation}
\end{subequations}
respectively, one can construct the hierarchy of conserved densities and fluxes along with functions $\Phi_1^{(k)}$. Specifically, relations (\ref{def-dens-flux}), which define recursively densities and fluxes, now simplify to
\begin{subequations} \label{clH1Q10}
\begin{equation}
\rho_{n,m}^{(k)}\,=\,{\cal{R}}_-^k \left(\rho_{n,m}^{(0)}\right)\,,\qquad \sigma_{n,m}^{(k)}\,=\,{\cal{R}}^k \left(\sigma_{n,m}^{(0)}\right)\,,
\end{equation}
where ${\cal{R}}_-$ is the ``negative'' part of $\cal{R}$, i.e.
\begin{equation}
{\cal{R}}_-\,=\,\sum_{\ell = -\infty}^{-1} \ell\, {\cal{S}}_n^{\ell}\left(R_1\right) \partial_{u_{n+\ell,m}}\,-\,\partial_\alpha\,.
\end{equation}
In terms of the above densities, functions $\Phi_1^{(i)}$ can be expressed as
\begin{equation}
\Phi_1^{(i)} \,=\, R_1 \partial_{u_{n,m}} \rho_{n-1,m}^{(i)}\,, \quad i=0,1,2,\ldots \,. \label{H1Q1F1k}
\end{equation}
\end{subequations}

The hierarchy of symmetries is given by
\begin{subequations} \label{sH1Q10}
\begin{equation}
V_k \,=\, R_k\,\partial_{u_{n,m}} + \sum_i \Phi_k^{(i)} \partial_{\phi^{(i)}}\,,\quad k=1,2,3,\ldots\,,
\end{equation}
where $\Phi_1^{(i)}$ are defined in (\ref{H1Q1F1k}) and, in terms of which, $\Phi_k^{(i)}$ are determined by relation (\ref{Fkl}) which now becomes
\begin{equation}
\Phi_k^{(i)} \,=\,\sum_{j=0}^{k-1} \left(\begin{array}{c} k-1 \\ j\end{array} \right) (-1)^{k+j-1} {\cal{R}}^{j}\left( R_1 \partial_{u_{n,m}} \rho_{n-1,m}^{(k+i-j-1)}\right)\,.
\end{equation}
\end{subequations}

On the other hand, the hierarchy of master symmetries is given by
\begin{subequations} \label{msH1Q10}
\begin{equation}
G_1 \,=\, n R_1 \partial_{u_{n,m}}\,+\,\sum_i \left(n \Phi_1^{(i)}+\phi^{(i+1)} \right)\partial_{\phi^{(i)}} - \partial_\alpha
\end{equation}
and
\begin{eqnarray}
G_k &=&\left(\frac{n}{k-1}\, R_k - \sum_{j=0}^{k-2} \left(\begin{array}{c} k-2 \\ j\end{array} \right) R_{j+1} \phi^{(k-j-1)}\right)\,\partial_{u_{n,m}}\nonumber\\
&&+ \sum_i\left\{ \frac{n}{k-1}\,\Phi_k^{(i)}+ \frac{i!}{(k+i-1)!}\phi^{(k+i)} - \sum_{j=0}^{k-2}  \left(\begin{array}{c} k-2 \\ j\end{array} \right) \Phi^{(i)}_{j+1} \phi^{(k-j-1)} +  \right.\nonumber \\
&&\left.\qquad \qquad \sum_{j=0}^{k-2} \frac{(-1)^{k+j}}{i + k - j - 1} \left(\begin{array}{c} k-2 \\ j\end{array} \right)  {\cal{R}}^j\left(\Phi_1^{(i+k-j-1)} \right) \right\} \,\partial_{\phi^{(i)}}\nonumber \\
&& +\,\, \xi_k(\alpha)\,\partial_\alpha\,+\,\zeta_k(\alpha,\beta)\,\partial_\beta ,\qquad  k =2,3,\ldots\,,
\end{eqnarray}
with
\begin{eqnarray}
{\mbox{for H1}} &:& \xi_k = 0\,,\qquad\qquad\quad \zeta_k = \frac{(k-2)!}{(\alpha-\beta)^{k-1}}\,,\\
{\mbox{for Q1}}_0 &:&  \xi_k = \frac{-(k-2)!}{\alpha^{k-1}}\,,\quad \zeta_k =  \frac{(k-2)!}{(\alpha-\beta)^{k-1}}-\frac{(k-1)!}{\alpha^{k-1}}\,.
\end{eqnarray}
\end{subequations}

Computing the commutators of the above symmetries, one is led to the following relations.
$$ \begin{array}{ll}
&[ V_i,V_j ] \, = \,  0\,, \\
&[ G_1, V_j ]  \,=\, V_{j+1}\,,\quad [G_k , V_j] \,=\, \frac{(k-2)! \,j!}{(k+j-1)!} V_{k+j}\,,\quad k = 2,3,\ldots\,,\\
&[G_1,G_j] \,=\,G_{j+1}\,,\quad  [G_i,G_j] \,=\, (j-i) \frac{(i-2)!\,(j-2)!}{(i+j-2)!}\,G_{i+j}\,,\quad i,j \,>\, 1\,.
\end{array}$$
These suggest to rescale the symmetry generators as
\begin{equation}
\begin{array}{ll}
& {\cal{V}}_k\,=\,\frac{1}{(k-1)!} V_k\,,\quad k=1,2,\ldots\,,\\
& {\cal{G}}_1\,=\,G_1\,,\quad {\cal{G}}_k\,=\,\frac{1}{(k-2)!} G_k\,,\,\,\, k>1\,,
\end{array} \label{rescVG}
\end{equation}
in terms of which the above commutation relations become
\begin{eqnarray}
&& [ {\cal{V}}_i,{\cal{V}}_j ] \, = \,  0 \,,\nonumber \\
&& [ {\cal{G}}_i, {\cal{V}}_j ]  \,=\, j\,{\cal{V}}_{j+i} \,, \label{cVa}\\
&& [{\cal{G}}_i,{\cal{G}}_j] \,=\, (j-i) \,{\cal{G}}_{i+j}\,. \nonumber
\end{eqnarray}
These relations imply that the algebra $\mathfrak{n}$ of the above symmetry generators is the semi-direct sum of the abelian algebra $\{{\cal{V}}_k\}$ and the centerless Virasoro algebra $\{{\cal{G}}_k\}$, ${\mathfrak{n}} = \{{\cal{V}}_k\} \oplus_s \{{\cal{G}}_k\}$. As consequence of the covariance of equations H1 and Q1$_0$, the symmetry generators $\tilde{\cal{V}}_k$, $\tilde{\cal{G}}_k$ will satisfy the same commutation relations and, hence, will form the same algebra. \\

\noindent {\sl{Remark}} \hspace{.15cm} Working out explicitly some of the commutators for equations H2, Q1$_\delta$ and Q2, it is apparent that their symmetries satisfy similar commutation relations suggesting that they form, after a suitable rescaling, similar algebras. However general expressions, like (\ref{clH1Q10}-\ref{msH1Q10}), are difficult to be derived in terms of which this statement can be proven. \hfill $\Box$ \\

\section{The symmetry algebra of H1} \label{symH1}

The symmetry algebra $\mathfrak{h}$ of H1 is generated by its point and generalized symmetries. The first ones have been found in \cite{RHs,TTX} and the generalized symmetries were the subject of the previous section. Here, we will explicitly compute the commutators among the generators of these three sub-algebras and, consequently, find the structure constants of $\mathfrak{h}$.

Let us begin with the sub-algebra $\mathfrak{p}$ of point symmetries generated by the vector fields \cite{RHs,TTX}
\begin{eqnarray*}
&& {\cal{T}}_1\,=\,\partial_{u_{n,m}}\,,\quad {\cal{T}}_2\,=\,(-1)^{n+m}\partial_{u_{n,m}}\,,\quad {\cal{V}}_0\,=\,(-1)^{n+m}u_{n,m}\partial_{u_{n,m}}\,,\\
&& {\cal{G}}_0\,=\,-\,\frac{u_{n,m}}{2}\partial_{u_{n,m}}\,-\,\alpha\partial_\alpha-\beta \partial_\beta + \sum_{j \ge 1} j \phi^{(j)} + \sum_{j \ge 1} j \tilde{\phi}^{(j)} \partial_{\tilde{\phi}^{(j)}},\quad {\cal{C}}\,=\,\partial_\alpha + \partial_\beta\,.\label{tssH1}
\end{eqnarray*}
One easily verifies that the only non-vanishing commutators among them are the following ones.
$$[{\cal{T}}_1,{\cal{V}}_0]\,=\,{\cal{T}}_2\,,\quad [{\cal{T}}_2,{\cal{V}}_0]\,=\,{\cal{T}}_1\,, \quad [{\cal{G}}_0,{\cal{T}}_1] = \frac{1}{2} {\cal{T}}_1\,,\quad [{\cal{G}}_0,{\cal{T}}_2] = \frac{1}{2} {\cal{T}}_2\,.$$

\noindent{\sl{Remark}} \hspace{.15cm} The reason for denoting two of the above generators by ${\cal{V}}_0$ and ${\cal{G}}_0$ is that these two specific symmetries are related to the hierarchies of symmetries ${\cal{V}}_i$, $\tilde{\cal{V}}_i$ and master symmetries ${\cal{G}}_i$, $\tilde{\cal{G}}_i$, respectively. As it was noted in Section \ref{secMS}, ${\cal{G}}_2$ can be derived from ${\cal{G}}_1$ by applying to the latter the recursion operator which was given in \cite{mwx}, namely operator $R_1 ({\cal{S}}_n+1) \Delta_n^{-1} R_1 ({\cal{S}}_n-{\cal{S}}_n^{-1})$. In exactly the same fashion, ${\cal{G}}_1$ can be constructed from ${\cal{G}}_0$ by applying to it the same recursion operator and then using the arguments mentioned in Section \ref{secMS}. Hence, the scaling generator ${\cal{G}}_0$ is the seed symmetry from which the whole hierarchy of master symmetries is derived by applying successively to it the pseudo-difference recursion operator $R_1 ({\cal{S}}_n+1) \Delta_n^{-1} R_1 ({\cal{S}}_n-{\cal{S}}_n^{-1})$. Similarly, the result of the application of the same operator on ${\cal{V}}_0$ is the first symmetry ${\cal{V}}_1$ implying that the former can be taken as the seed symmetry of the hierarchy ${\cal{V}}_i$. \hfill $\Box$ \\

Before considering the commutators among generators of the three sub-algebras, let us rescale the generalized symmetries and re-define them as follows.
\begin{equation}
\begin{array}{l}
{\cal{W}}_k\,:=\,(\alpha-\beta)^k\,{\cal{V}}_k\,,\quad {\cal{H}}_k\,:=\,(\alpha-\beta)^k\,{\cal{G}}_k\,, \\
\tilde{\cal{W}}_k\,:=\,(\beta-\alpha)^k\,\tilde{\cal{V}}_k\,,\quad \tilde{\cal{H}}_k\,:=\,(\beta-\alpha)^k\,\tilde{\cal{G}}_k\,.
\end{array} \label{WHdef}
\end{equation}
It should be noted that since $\alpha$ and $\beta$ are parameters, we can freely multiply any symmetry by a function of the lattice parameters with the resulting vector field being again symmetry generator of H1. For instance, the infinitesimal criterion for ${\cal{H}}_k$ to be a symmetry of H1 is the same with the one for ${\cal{G}}_k$ differing by an overall multiplicative factor $(\alpha-\beta)^k$.
The commutation relations among the generators ${\cal{W}}_k$, ${\cal{H}}_k$ of ${\mathfrak{n}}$ and the corresponding ones of $\mathfrak{m}$ follow from (\ref{cVa}), (\ref{WHdef}) and read as follows.
\begin{eqnarray}
&& [ {\cal{W}}_i,{\cal{W}}_j ] = 0 \phantom{(j-i) {\cal{H}}_{i+j}+ i\,{\cal{H}}_{i}- j {\cal{H}}_{j}} \qquad \,\,\, [ \tilde{\cal{W}}_i,\tilde{\cal{W}}_j ]  = 0 \nonumber \\
&& [{\cal{H}}_i,{\cal{W}}_j] = j\left({\cal{W}}_{i+j}-{\cal{W}}_j\right)\qquad \qquad \qquad\qquad \! [\tilde{\cal{H}}_i, \tilde{\cal{W}}_j ]  = j \left(\tilde{\cal{W}}_{i+j}-\tilde{\cal{W}}_j\right)   \label{cWa}\\
&& [{\cal{H}}_i,{\cal{H}}_j] = (j-i) {\cal{H}}_{i+j}+ i\,{\cal{H}}_{i}- j {\cal{H}}_{j} \qquad \quad \!\! [\tilde{\cal{H}}_i,\tilde{\cal{H}}_j] = (j-i) \tilde{\cal{H}}_{i+j} + i \tilde{\cal{H}}_{i} - j \tilde{\cal{H}}_{j} \nonumber
\end{eqnarray}

It is straightforward to verify that all the point symmetries commute\footnote{Although all the other point symmetries commute with the hierarchies of ${\cal{V}}_i$ and ${\cal{G}}_i$, scaling symmetry ${\cal{G}}_0$ satisfies $[{\cal{G}}_0,{\cal{V}}_j]\,=\,j\,{\cal{V}}_j$ and $[{\cal{G}}_0,{\cal{G}}_j]\,=\,j\,{\cal{G}}_j$. From the latter and definitions (\ref{WHdef}) follows that ${\cal{G}}_0$ commutes with ${\cal{W}}_i$ and ${\cal{H}}_i$.} with the above generalized symmetries, i.e. $[{\mathfrak{p}}, {\mathfrak{n}}] = [{\mathfrak{p}}, {\mathfrak{m}}]=0$. Thus, symmetry algebra $\mathfrak{h}$ of H1 is generated by the sub-algebra $\mathfrak{p}$ of point symmetries and the sub-algebra $\mathfrak{g}$ of generalized symmetries, and, in particular, ${\mathfrak{h}} = {\mathfrak{p}} \oplus {\mathfrak{g}}$. For the algebra ${\mathfrak{g}}$, we know that ${\mathfrak{g}} = {\mathfrak{n}} +{\mathfrak{m}}$ as a vector space. Now, we will try to find the commutators among the generators of these two sub-algebras.

For this purpose, we will extend the action of ${\cal{V}}_1$, ${\cal{V}}_2$,  and ${\cal{G}}_1$, ${\cal{G}}_2$, in the direction of the potential $\tilde{\phi}^{(1)}$. Once these actions are found, the covariance of H1 implies the corresponding actions of $\tilde{{\cal{V}}}_1$, $\tilde{{\cal{V}}}_2$, $\tilde{{\cal{G}}}_1$ and $\tilde{{\cal{G}}}_2$ on $\phi^{(1)}$. Finally, these results combined with definitions (\ref{WHdef}) will deliver the corresponding action of ${\cal{W}}_i$, $\tilde{\cal{W}}_i$, ${\cal{H}}_i$ and $\tilde{\cal{H}}_i$ for $i=1,2$.

The action of ${\cal{V}}_1$ and ${\cal{G}}_1$ on $\tilde{\phi}^{(1)}$ is given by
$$\tilde{\Phi}_1^{(1)}\,\partial_{\tilde{\phi}^{(1)}}\,:=\,\frac{1}{\alpha-\beta}\left(\frac{2}{w_{n,m}}\left(\frac{1}{v_{n,m-1}}\,+\,\frac{u_{n,m-1}-u_{n-1,m}}{\alpha-\beta}\right)\,-\,\frac{1}{\alpha-\beta}\right)\partial_{\tilde{\phi}^{(1)}}$$
and $n\,\tilde{\Phi}_1^{(1)}\,\partial_{\tilde{\phi}^{(1)}}$, respectively. In the above relation, $w_{n,m}$ is the function defined in (\ref{defw}) and
$$v_{n,m}\,:=\,u_{n,m+1}-u_{n,m-1}\,. $$
The action of ${\cal{V}}_2$ on ${\tilde{\phi}^{(1)}}$ can be derived from the definition of the latter, ${\cal{V}}_2=[{\cal{G}}_1,{\cal{V}}_1]$, and it is 
$$\tilde{\Phi}_2^{(1)}\,\partial_{\tilde{\phi}^{(1)}} \,:= \, {\cal{R}}\left(\tilde{\Phi}_1^{(1)}\right)\,\partial_{\tilde{\phi}^{(1)}}\,.$$
Finally, ${\cal{G}}_2$ acts on $\tilde{\phi}^{(1)}$ as
$$\left(n \tilde{\Phi}_2^{(1)} - \tilde{\Phi}_1^{(1)} \phi^{(1)} - \frac{\tilde{\phi}^{(1)} + \rho_{n,m}^{(1)} + \frac{2}{w_{n,m}} \left(\frac{1}{v_{n,m-1}}+\frac{2}{u_{n-1,m-1}-u_{n,m}}\right)}{(\alpha-\beta)^2}-\frac{m}{(\alpha-\beta)^3} \right) \partial_{\tilde{\phi}^{(1)}}\,.$$

Now, the commutation relations
\begin{subequations} \label{crnm}
\begin{equation}
[{\cal{H}}_1,\tilde{{\cal{H}}}_1]\,=\,{\cal{H}}_1 - \tilde{{\cal{H}}}_1\,,\quad [{\cal{H}}_1,\tilde{{\cal{H}}}_2] \,=\,2 \left({\cal{H}}_1- \tilde{{\cal{H}}}_2\right)
\end{equation}
and
\begin{equation}
[{\cal{H}}_2,\tilde{{\cal{W}}}_1]= -\tilde{{\cal{W}}}_1,\quad  [{\cal{H}}_2,\tilde{{\cal{H}}}_2] = 4 \left({\cal{H}}_2-\tilde{{\cal{H}}}_2\right)+3 \left({\cal{H}}_1-\tilde{{\cal{H}}}_1\right)
\end{equation}
\end{subequations}
can easily be verified. From commutation relations (\ref{cWa}), (\ref{crnm}) and the Jacobi identity it follows that 
\begin{subequations} \label{cWa2}
\begin{eqnarray}
&& [{\cal{W}}_k,\tilde{{\cal{W}}}_\ell]\,=\,0\,, \nonumber \\
&& [{\cal{H}}_k,\tilde{{\cal{W}}}_\ell]\,=\,-\ell\,\left( k \,\tilde{{\cal{W}}}_\ell \,+\,\sum_{j=1}^{\ell-1}\,C_{k,\ell}^j\, \tilde{\cal{W}}_j\right)\,,\nonumber \\
&& [\tilde{\cal{H}}_k,{{\cal{W}}}_\ell]\,=\,-\,\ell\,\left(k\,{{\cal{W}}}_\ell \,+\,\sum_{j=1}^{\ell-1}\,C_{k,\ell}^j\, {\cal{W}}_j\right)\,,  \\
&& [{\cal{H}}_k,\tilde{{\cal{H}}}_\ell]\,=\,k\,\ell\,\left({\cal{H}}_k-\tilde{\cal{H}}_\ell\right)\,+\, \sum_{j=1}^{k-1} (2k-j)\,C_{\ell ,k}^j\, {\cal{H}}_j \,-\, \sum_{j=1}^{\ell-1} (2 \ell-j)\,C_{k,\ell}^{j}\, \tilde{\cal{H}}_j \,, \nonumber
\end{eqnarray}
where
\begin{equation}
C_{k,\ell}^j\,:=\, \left(\begin{array}{c} k+\ell-j-1 \\ \ell-j+1 \end{array} \right)\,. 
\end{equation}
\end{subequations}
The above commutators hold for any $k, \ell = 1,2,3,\ldots$, and along with the ones in (\ref{cWa}) describe the sub-algebra $\mathfrak{g}$ of the generalized symmetries of equation H1.

\section{Symmetries and differential -- difference equations} \label{symdd}

The relation among symmetries of the ABS equations and differential-difference equations of Volterra type is well known \cite{LPSY,mwx1}. In particular, the former can be regarded as particular cases of Yamilov's discrete Krichever-Novikov (YdKN) equation \cite{Y1983,Y2006}. It is also known that YdKN equation admits a master symmetry in terms of which the whole hierarchy of symmetries can be constructed \cite{SY}. In this section, we present for the particular cases of the YdKN equation corresponding to symmetries of H1-H3 and Q1-Q3, the hierarchy of master symmetries along with their Lax pairs and discuss the algebraic properties of latter \cite{FM,MF}.

In this interpretation, we assume that $u$ also depends on an infinite number of time variables $t_k$ and $\tau_k$, which essentially correspond to the group parameters of the transformations generated by the symmetry generators $V_k$ and $G_k$, respectively. The symmetry condition now translates to the requirement the flows
\begin{equation}
\frac{\partial u_{n,m}}{\partial t_k}\,=\,R_k \label{ddh1}
\end{equation}
to commute with the discrete equations under consideration, i.e. $\partial_{t_k} Q_{n,m}=0$.
The latter relation holds on solutions of $Q_{n,m}=0$ by taking into account (\ref{ddh1}). Moreover, from the construction in Section \ref{cl-ms}, it follows that the relations \begin{equation}
\partial_{t_k}\rho_{n,m}^{(i)}\,=\,\Delta_n \Phi_k^{(i)}\,,\quad k=1,2,\ldots,\,\quad i=0,1,\ldots, \label{cldd}
\end{equation}
hold on solutions of (\ref{ddh1}). In other words, they are conservation laws of the corresponding differential--difference equations (\ref{ddh1}). Actually, for $k=1$ and up to rescaling, the first two conservation laws, which correspond to $i=0$ and $i=1$, respectively, coincide with the first two integrability conditions appearing in the classification of integrable Volterra type equations, cf. relations (A.11) in \cite{ASY}. On the other hand, the third conservation law, i.e. for $i=2$, is equivalent to the third integrability condition given in \cite{ASY}.

\begin{subequations} \label{ddh2}
In the same fashion, for master symmetries $G_k$ corresponding to H1-H2 and Q1-Q2, we have that the flows
\begin{eqnarray}
&& \frac{\partial u_{n,m}}{\partial \tau_1}\,=\, n\,R_1 \,, \label{ddh2a}\\
&& \frac{\partial u_{n,m}}{\partial \tau_\ell}\,=\,\frac{n}{\ell-1} R_\ell - \sum_{j=0}^{\ell-2} \left(\begin{array}{c} \ell-2 \\ j\end{array} \right) R_{j+1} \phi^{(\ell-j-1)} - {\cal{R}}^{\ell-2}\left(\frac{r_{n,m}}{2 \, w_{n-1,m}}\right)\,,\quad \ell \ge 2\,, \label{ddh2b}
\end{eqnarray}
accompanied with
\begin{equation}
\frac{\partial \alpha}{\partial \tau_k}\,=\,\xi_k(\alpha)\,,\quad \frac{\partial \beta}{\partial \tau_k}\,=\,\zeta_k(\alpha,\beta)\,,\quad k=1,2,\ldots, \label{ddh2c}
\end{equation}
\end{subequations}
commute with the corresponding discrete equation, i.e. $\partial_{\tau_k} Q_{n,m}=0$ holds on solutions of  $Q_{n,m}=0$ in view of (\ref{ddh2}). For equations H3 and Q3 similar considerations hold using the corresponding characteristics of master symmetries (\ref{msH3Q3}).

Concentrating on the hierarchies of differential-difference equations (\ref{ddh1}) and their master symmetries (\ref{ddh2}), we will see how the former possess an isospectral zero curvature representation while the latter possess the corresponding non-isospectral one \cite{FM,MF}. In other words, we will construct Lax pairs for all these equations in which the spectral parameter, the role of which is played now by the lattice parameter $\beta$, will be considered as a function of the time variables $\tau_k$ and independent of $t_k$. Specifically, the second relation of (\ref{ddh2c}) will provide us with the particular dependence of the spectral parameter $\beta$ on the time variables $\tau_k$.

To start with, let us construct the spectral problem to which these hierarchies are related to. It has the form
\begin{subequations}\label{Lpdp}
\begin{equation}
\Psi_{n+1,m}\,=\,{\rm{L}}\,\Psi_{n,m}\,,
\end{equation}
where
\begin{equation}
{\rm{L}}\,=\,\frac{1}{\sqrt{\kappa(\alpha,\beta) f_{n,m}}} \left(\begin{array}{cc} \partial_{u_{n+1,m+1}} Q_{n,m} & -\partial_{u_{n+1,m+1}} \partial_{u_{n,m+1}} Q_{n,m}\\ Q_{n,m} & -\partial_{u_{n,m+1}} Q_{n,m} \end{array} \right) \,.
\end{equation}
\end{subequations}
Here, $Q_{n,m}$ and its derivatives are evaluated at $u_{n,m+1}=u_{n+1,m+1}=0$ and $\beta$ plays the role of the spectral parameter. Matrix $\rm{L}$ is coming from the corresponding Lax pair of the ABS equation to which equations (\ref{ddh1}), (\ref{ddh2}) are related to. In particular, matrix ${\rm{L}}$ follows from the multidimensional consistency of the corresponding ABS equation $Q_{n,m}=0$ \cite{BS,N2002} and the above form of matrix $\rm{L}$ was given in \cite{X}.

The isospectral hierarchy can be derived systematically and has the following form \cite{X}
\begin{subequations} \label{Lpcp}
\begin{equation}
\frac{\partial \Psi_{n,m}}{\partial t_k}\,=\,{\rm{M}}_k\,\Psi_{n,m}\,,\quad k=1,2,\ldots. 
\end{equation}
The matrices ${\rm{M}}_k$ are given recursively by
\begin{equation}
{\rm{M}}_k\,=\,{\cal{R}}^{\,k-1}{\rm{M}}_1\,,\quad k=1,2,\ldots,
\end{equation}
with
\begin{equation}
{\rm{M}}_1\,:=\,\left(\frac{1}{w_{n,m}}\,-\,\frac{1}{2}\,\partial_{u_{n+1,m}}\right)\left(-f_{n,m} \,{\rm{L}}^{-1} \partial_{u_{n,m}} {\rm{L}}\right)\,.
\end{equation}
\end{subequations}
The compatibility condition $\partial_{t_k} {\rm{L}}\, + {\rm{L}}{\rm{M}}_k\,=\,\left( {\cal{S}}_n {\rm{M}}_k\right) {\rm{L}}$ of (\ref{Lpdp}), (\ref{Lpcp}) is satisfied in view of equation (\ref{ddh1}).

On the other hand, the non-isospectral hierarchy follows the pattern of master symmetries with $R_k$ being replaced by matrix ${\rm{M}}_k$ and the last term involving polynomial $r_{n,m}$ being substituted by a suitable matrix $\rm{H}$. More precisely, the non-isospectral hierarchy corresponding to (\ref{ddh2}) is given by
\begin{subequations} \label{Lpcpg}
\begin{eqnarray}
\frac{\partial \Psi_{n,m}}{\partial \tau_1} &=& {\rm{N}}_1\,\Psi_{n,m}\,:=\,\left( n\,{\rm{M}}_1\right)\,\Psi_{n,m}\,, \label{Lpcpg1} \\
\frac{\partial \Psi_{n,m}}{\partial \tau_\ell} &=&{\rm{N}}_\ell\,\Psi_{n,m} := \left(\frac{n}{\ell-1}{\rm{M}}_\ell - \sum_{j=0}^{\ell-2} \left(\begin{array}{c} \ell-2 \\ j\end{array} \right)  \phi^{(\ell-j-1)} {\rm{M}}_{j+1} - {\cal{R}}^{\,\ell-2}\left({\rm{H}}\right) \right) \Psi_{n,m}, \,\,\ell \ge 2, \label{Lpcpg4} 
\end{eqnarray}
along with
\begin{equation}
\frac{\partial \alpha}{\partial \tau_\ell}\,=\,\xi_\ell(\alpha)\,,\quad \frac{\partial \beta}{\partial \tau_\ell}\,=\,\zeta_\ell(\alpha,\beta)\,,\quad \ell=1,2,3,\ldots. \label{Lpcpgp}
\end{equation}
\end{subequations}
On the other hand, for H3 and Q3 in particular, we have
\begin{eqnarray}
\frac{\partial \Psi_{n,m}}{\partial \tau_3} &=& \left(\frac{n}{2}\,{\rm{M}}_3-\frac{n}{12}\,{\rm{M}}_1-\phi^{(1)}\,{\rm{M}}_2-\phi^{(2)}\,{\rm{M}}_1-{\cal{R}}\,{\rm{H}}\right) \Psi_{n,m} \label{Lpcpg3}\,,\\
\frac{\partial \Psi_{n,m}}{\partial \tau_6} &=& \left(\frac{n}{5} {\rm{M}}_6 - \frac{n}{30} {\rm{M}}_2 - \sum_{j=0}^{4} \left(\begin{array}{c} 4 \\ j\end{array} \right) \phi^{(5-j)} {\rm{M}}_{j+1} -{\cal{R}}^{\,4} \left({\rm{H}}\right)\right)\Psi_{n,m} \,,\\
\frac{\partial \Psi_{n,m}}{\partial \tau_7} &=& \left(\frac{n}{6} {\rm{M}}_7 - \frac{n}{12} {\rm{M}}_3 - \sum_{j=0}^{5} \left(\begin{array}{c} 5 \\ j\end{array} \right) \phi^{(6-j)} {\rm{M}}_{j+1}- {\cal{R}}^{\,5} \left({\rm{H}}\right)\right)\Psi_{n,m}\,.
\end{eqnarray}
Matrix $\rm{H}$ appearing in the above relations can be derived, up to a point, in a systematic way and its final form depends on the particular equation. These precise forms are given in the following list.
\begin{subequations}
\begin{eqnarray}
{\mbox{H1, H2, Q1 }} &:& {\rm{H}}\,:=\,\rho_{n,m}^{(1)} {\rm{M}}_1\,-\,2\,\left({\cal{S}}_n^{-1} R_1\right)\partial_{u_{n-1,m}} {\rm{M}}_1 - \partial_\alpha {\rm{M}}_1 \\
{\mbox{H3, Q3}} &:& {\rm{H}}\,:=\,\rho_{n,m}^{(1)} {\rm{M}}_1\,-\,2\,\left({\cal{S}}_n^{-1} R_1\right)\partial_{u_{n-1,m}} {\rm{M}}_1  + \frac{\alpha}{2}\,\partial_\alpha {\rm{M}}_1 + \left( \begin{array}{cc} 0 & 0 \\ 0 & \frac{1}{4}\end{array} \right)\\
{\mbox{Q2 }} &:& {\rm{H}}\,:=\,\rho_{n,m}^{(1)} {\rm{M}}_1\,-\,2\,\left({\cal{S}}_n^{-1} R_1\right)\partial_{u_{n-1,m}} {\rm{M}}_1 - \partial_\alpha {\rm{M}}_1 + \left( \begin{array}{cc} 0 & 0 \\ -2 & 0 \end{array} \right)
\end{eqnarray}
\end{subequations}
The compatibility condition $\partial_{\tau_k} {\rm{L}}\, + {\rm{L}}{\rm{N}}_k\,=\,\left( {\cal{S}}_n {\rm{N}}_k\right) {\rm{L}}$ of (\ref{Lpdp}), (\ref{Lpcpg}) is satisfied in view of equation (\ref{ddh2}).\\

\noindent{\sl{Remark}} \hspace{.15cm} We would like to emphasize that the formulae for the hierarchies of differential-difference equations (\ref{ddh1}), their canonical conservation laws (\ref{cldd}), as well as their Lax pairs (\ref{Lpdp}), (\ref{Lpcp}), are valid for the symmetries corresponding to any of the ABS equations, i.e. including Q4 as well. However, only the formulae for the first master symmetry (\ref{ddh2a}) and its Lax pair (\ref{Lpcpg1}) are also valid for Q4 \cite{X}. \hfill $\Box$\\

As explained above, matrices $\rm{L}$, ${\rm{M}}_1$, ${\rm{N}}_1$ and ${\rm{N}}_2$ were derived in a systematic way. The remaining matrices follow from the consistency of the different linear equations for the potential $\Psi_{n,m}$. More precisely, the compatibility condition of equation (\ref{Lpcp}) with $k=1$, and equation (\ref{Lpcpg1}) yields ${\rm{M}}_2$, while the compatibility condition of (\ref{Lpcp}) with $k=2$, and (\ref{Lpcpg1}) is matrix ${\rm{M}}_3$. Thus, the matrices ${\rm{M}}_k$ of the isospectral hierarchy are constructed in the same way hierarchy (\ref{ddh1}) was derived.
$${\rm{M}}_{k+1}\,=\, \llbracket {\rm{M}}_k,{\rm{N}}_1\rrbracket\,:=\,\frac{\partial {\rm{M}}_k}{\partial \tau_1}\,-\,\frac{\partial {\rm{N}}_1}{\partial t_k} \,+\,\left[{\rm{M}}_k\,,\,{\rm{N}}_1 \right]\,,\quad k =1,2,\ldots\,.$$
Similarly, the non-isospectral hierarchy can be generated in the same manner starting with equations (\ref{Lpcpg1}) and (\ref{Lpcpg4}) with $\ell=2$ and their consistency. This approach leads to 
$${\rm{N}}_{k+1}\,=\, \llbracket {\rm{N}}_k,{\rm{N}}_1\rrbracket\,:=\,\frac{\partial{\rm{N}}_k}{\partial \tau_1}-\frac{\partial{\rm{N}}_1}{\partial \tau_k}+ \left[\,{\rm{N}}_k, {\rm{N}}_1\right]\,,\quad k=2,3,\ldots\,.$$
Finally, the commutativity of hierarchy (\ref{ddh1}) implies that
$$\llbracket {\rm{M}}_k,{\rm{M}}_\ell\rrbracket\,:=\,\frac{\partial {\rm{M}}_k}{\partial t_\ell}\,-\, \frac{\partial {\rm{M}}_\ell}{\partial t_k} \,+\,\left[{\rm{M}}_k\,,\,{\rm{M}}_\ell \right]\,=\,0\,.$$
In all the above relations the derivative with respect to $\tau_k$ is understood as
$$\frac{\partial \phantom{\tau_k}}{\partial \tau_k}\,=\,\sum_{i=-\infty}^{\infty} \frac{\partial u_{n+i,m}}{\partial \tau_k}\frac{\partial \phantom{u_{n+i,m}}}{\partial u_{n+i,m}}\,+\,\xi_k(\alpha)\,\partial_\alpha\,+\,\zeta_k(\alpha,\beta)\,\partial_\beta\,,\quad k=1,2,\ldots. $$

The bracket $\llbracket \cdot , \cdot \rrbracket$ was introduced in \cite{MF} where the general algebraic structure related to discrete zero curvature equations was developed and was illustrated by the examples of Volterra and Toda lattice hierarchies. In particular, in those examples it was shown that the Lax operators form Virasoro type algebras. In fact, employing the analysis and the results of Section \ref{secH1Q1}, it can be verified that the following relations hold for the hierarchies of Lax operators corresponding to equations H1 and Q1$_0$.
\begin{equation}
\llbracket {\rm{M}}_i,{\rm{M}}_j\rrbracket=0,\quad  \llbracket {\rm{M}}_j,{\rm{N}}_i\rrbracket=j\,{\rm{M}}_{i+j}, \quad \llbracket {\rm{N}}_j,{\rm{N}}_i\rrbracket= (j-i) {\rm{N}}_{i+j} . \label{Lax-algebra}
\end{equation}
For equations H2, H3 and Q1$_\delta$-Q3, we remark that one could employ the results of \cite{MF} and use the fact \cite{Y2006} that, under some easily verified conditions, the first member of hierarchy (\ref{ddh1})  is related through two different Miura transformations to either the Volterra equation $\partial_t u_n = u_n (u_{n+1}-u_{n-1})$, cf. also \cite{LPSY}, or the Toda system $\partial_t u_n = u_n (v_{n+1}-v_n)$, $\partial_t v_n = u_n - u_{n-1}$.

\section{Conclusions and perspectives} \label{conper}

We have presented a constructive approach to the derivation of symmetries and canonical conservation laws for the integrable difference equations of the ABS classification. The basic ingredients  in this construction are the first symmetry $V_1$, the first master symmetry $G_1$ and the first canonical conservation law $\left(\rho^{(0)}_{n,m},\sigma^{(0)}_{n,m}\right)$ in terms of which the rest members in these hierarchies can be derived systematically. The approach for the construction of canonical conservation laws can be applied to any equation and it would be interesting to apply this method to the non-autonomous deformed H equations \cite{ABS1} the symmetries of which were studied in \cite{XP}, as well as to the most general equation in the class of Kleinian invariant equations, i.e. equation Q4 with arbitrary parameters \cite{V}. 

For equations H1--H3, Q1--Q3 the second master symmetry $G_2$ was given explicitly and it was used with $G_1$ in the construction of a corresponding chain of master symmetries. All the higher order master symmetries, including $G_2$, are non local since their characteristics depend explicitly on the potentials $\phi^{(i)}$ corresponding to canonical conservation laws. It was proven that these non-local master symmetries map local symmetries to local symmetries. In particular, for equations H1 and Q1$_0$, it was shown that their symmetries and master symmetries in each lattice direction form centerless Virasoro type algebras, and the whole symmetry algebra for H1 was derived. The construction of the second master symmetry of Q4 is still an open question. An interesting question related to master symmetries and conservation laws is if these master symmetries can be regarded as variational symmetries of the ABS equations, a Lagrangian of which is given in \cite{XNL}, and if these canonical conservation laws can be derived by implementation of Noether's theorem.

The interpretation of symmetries as differential-difference equations led us to the derivation of hierarchies of integrable differential-difference equations along with corresponding hierarchies of master symmetries and canonical conservation laws. We also presented isospectral and non-isospectral hierarchies to which these differential-difference equations and their master symmetries correspond to. It would be interesting to derive similar structures for the general YdKN equation, i.e. the first generalized symmetry of Adler's equation, which is not related to Volterra or Toda systems \cite{LPSY,Y2006}.

\section*{Acknowledgments}

I would particularly like to express my gratitude to Peter Olver, Artur Sergyeyev and Jing Ping Wang for useful discussions about master symmetries and conservations laws, and especially to Artur Sergyeyev for bringing to my attention references \cite{FF} and \cite{S}. I also wish to thank Anastasia Doikou, J{\"u}rgen Fuchs, Alexander Mikhailov, Frank Nijhoff, Vassilis Papageorgiou, Simon Ruijsenaars and Vitaly Tarasov for useful discussions about infinite dimensional algebras, especially in connection with the symmetry algebra of H1. This work was supported by the Newton International Fellowship grant NF082473 entitled ``Symmetries and integrability of lattice equations and related partial differential equations''.

\appendix
\section{The ABS equations and their polynomials}

With the identifications
$$u := u_{n,m},\quad x := u_{n+1,m},\quad y := u_{n,m+1},\quad z := u_{n+1,m+1},$$
the Adler-Bobenko-Suris equations are given in the following list.
\begin{eqnarray*} 
{\mbox{H1}} && (u-z)\, (x-y)\, -\,\alpha \,+ \, \beta \, = \,0 \\
{\mbox{H2}} & & (u-z)(x-y) +(\beta-\alpha) (u+x+y+z)  - \alpha^2 + \beta^2 = 0 \label{H2} \\
{\mbox{H3}} & & \alpha (u x+y z) - \beta (u y+x z) + \delta (\alpha^2-\beta^2) = 0 \label{H3} \\
{\mbox{Q1}} & & \alpha (u-y) (x- z) - \beta (u- x) (y -z)  + \delta^2 \alpha \beta (\alpha-\beta)= 0 \label{Q1}\\
{\mbox{Q2}} & & \alpha (u-y) (x- z) - \beta (u- x) (y -z) + \nonumber\\
 &&  \alpha \beta (\alpha-\beta) (u+x+y+z) - \alpha \beta (\alpha-\beta) (\alpha^2-\alpha \beta + \beta^2) = 0 \label{Q2}\\
{\mbox{Q3}} & & (\beta^2-\alpha^2) (u z+x y) + \beta (\alpha^2-1) (u x+y z) \nonumber\\
&&  - \alpha (\beta^2-1) (u y+x z) - \delta^2 (\alpha^2-\beta^2) (\alpha^2-1) (\beta^2-1)/(4 \alpha \beta)=0  \label{Q3} \\
{\mbox{Q4}} & & a_0 u x y z  + a_1 (u x y + x y z + y z u + z u x) + a_2 (u z + x y) \\
 & &+ \bar{a}_2 (u x+y z) + \tilde{a}_2 (u y+x z) + a_3 (u + x + y + z) + a_4 = 0\nonumber
\end{eqnarray*}
In Adler's parametrization \cite{A}, the parameters involved in Q4 are determined by the relations
\begin{eqnarray*}
&& a_0 = a+b \,,\,\,\,a_1=-a \beta - b \alpha\,,\,\,\,a_2=a \beta^2 + b \alpha^2\,, \\
&& \bar{a}_2 = \frac{a b (a+b)}{2 (\alpha-\beta)} + a \beta^2 - (2 \alpha^2 - \frac{g_2}{4}) b\,,\,\,
 \tilde{a}_2 = \frac{a b (a+b)}{2 (\beta-\alpha)} + b \alpha^2 - (2 \beta^2 - \frac{g_2}{4}) a\,, \\
&& a_3 = \frac{g_3}{2}a_0 - \frac{g_2}{4} a_1\,,\,\,\,a_4=\frac{g_2^2}{16}a_0-g_3 a_1\,,
\end{eqnarray*}
with 
$$a^2\, =\, p(\alpha)\,,\quad b^2 \,=\, p(\beta)\,,\quad p(x)\,=\,4 x^3-g_2 x - g_3\,.$$
In Hietarinta's parametrization \cite{H}, the parameters in Q4 have the following forms
\begin{eqnarray*}
&& a_0=a_4=k \,{\rm{sn}}(\alpha) \,{\rm{sn}}(\beta) \,{\rm{sn}}(\alpha-\beta),\quad a_1 = a_3 = 0,\\
&&a_2 ={\rm{sn}}(\beta-\alpha), \quad {\bar{a}}_2 = {\rm{sn}}(\alpha),\quad {\tilde{a}}_2 = -{\rm{sn}}(\beta),
\end{eqnarray*}
where ${\rm{sn}}$ denotes the Jacobi elliptic function ${\rm{sn}}(\alpha|k)$ with modulus $k$.

The polynomials related to the ABS equations are given in the following list.
\begin{itemize}
\item Equation H1
$$f_{n,m} = 1,\quad h(x,y,\alpha,\beta) = (x-y)^2,\quad \kappa(\alpha,\beta) = \beta -\alpha, \quad r_{n,m} = 0$$ 
\item Equation H2
\begin{eqnarray*}
&& f_{n,m}= 2(u_{n,m} + u_{n+1,m} + \alpha), \quad h(x,y,\alpha,\beta) = (x-y)^2 - (\alpha-\beta)^2,\\
&& \kappa(\alpha,\beta) = \beta -\alpha,\quad r_{n,m} = 4
\end{eqnarray*}
\item Equation H3
\begin{eqnarray*}
&& f_{n,m}= u_{n,m} u_{n+1,m} + \alpha \delta, \quad h(x,y,\alpha,\beta) = (\alpha y - \beta x) (\beta y -  \alpha x),\\
&& \kappa(\alpha,\beta) = \alpha^2 - \beta^2,\quad r_{n,m} = u_{n,m}^2
\end{eqnarray*}
\item Equation Q1
\begin{eqnarray*}
&& f_{n,m}= ((u_{n,m}-u_{n+1,m})^2 - \alpha^2 \delta^2)/\alpha, \\
&& h(x,y,\alpha,\beta) = \alpha \beta \left((x-y)^2 - (\alpha -\beta)^2 \delta^2\right),\\
&& \kappa(\alpha,\beta) = -\alpha \beta (\alpha-\beta) ,\quad r_{n,m} = 4\, \delta^2
\end{eqnarray*}
\item Equation Q2
\begin{eqnarray*}
&& f_{n,m}= ((u_{n,m}-u_{n+1,m})^2 - 2 \alpha^2 (u_{n,m}+u_{n+1,m}) + \alpha^4)/\alpha, \\
&& h(x,y,\alpha,\beta) =  \alpha \beta \left((x-y)^2 - 2 (\alpha -\beta)^2 (x+y) + (\alpha-\beta)^4\right),\\
&& \kappa(\alpha,\beta) = -\alpha \beta (\alpha-\beta) ,\quad r_{n,m} = 16\,u_{n,m}
\end{eqnarray*}
\item Equation Q3
\begin{eqnarray*}
&& f_{n,m}= \frac{-1}{4\alpha(\alpha^2-1)} (4 \alpha (u_{n+1,m}-\alpha u_{n,m}) (\alpha u_{n+1,m} - u_{n,m}) + (\alpha^2-1)^2 \delta^2), \\
&& h(x,y,\alpha,\beta) =  \frac{(\alpha^2-1) (\beta^2-1)}{4 \alpha \beta} \left(4 \alpha \beta (\alpha y-\beta x) (\beta y-\alpha x)+(\alpha^2-\beta^2)^2 \delta^2\right),\\
&& \kappa(\alpha,\beta) = (\alpha^2 -  \beta^2) (\alpha^2-1) (\beta^2 - 1) ,\quad r_{n,m} = u_{n,m}^2 -\delta^2
\end{eqnarray*}
\item Equation Q4 (Adler's parametrization)
\begin{eqnarray*}
&&f_{n,m}= \frac{(u_{n,m} u_{n+1,m}+ \alpha (u_{n,m}+u_{n+1,m}) + \frac{g_2}{4})^2 - (u_{n,m}+u_{n+1,m}+\alpha)(4 \alpha u_{n,m} u_{n+1,m}-g_3)}{a} , \\
&&\kappa(\alpha,\beta) = \frac{a b\left(a^2 b + a b^2 +  \left[12 \alpha \beta^2 - g_2 (\alpha+2\beta) - 3 g_3\right] a + \left[12 \beta \alpha^2 - g_2 (\beta + 2 \alpha) - 3 g_3\right] b \right)}{4 (\alpha-\beta)} ,\\
&& r_{n,m}\, =\, 4 u_{n,m}^3 -g_2 u_{n,m} - g_3
\end{eqnarray*}
Polynomial $h$ is omitted because of its length but it can be derived directly from its definition (\ref{defh}).
\end{itemize}

\end{document}